# Epileptic Seizure Forecasting: Probabilistic seizure-risk assessment and data-fusion

Nhan Duy Truong, Yikai Yang, Christina Maher, Armin Nikpour, and Omid Kavehei*

***Abstract*—Epileptic seizure forecasting, combined with the delivery of preventative therapies, holds the potential to greatly improve the quality of life for epilepsy patients and their caregivers. Forecasting seizures could prevent some potentially catastrophic consequences such as injury and death in addition to a long list of potential clinical benefits it may provide for patient care in hospitals. The challenge of seizure forecasting lies within the seemingly unpredictable transitions of brain dynamics into the ictal state. The main body of computational research on determining seizure risk has been focused solely on prediction algorithms, which involves a remarkable issue of balancing accuracy and false-alarms. In this paper, we developed a seizure-risk warning system that employs Bayesian convolutional neural network (BCNN) to provide meaningful information to the patient and provide a greater opportunity for him/her to be potentially more in charge of his/her health. We use scalp electroencephalogram (EEG) signals and release information on the certainty of our automatic seizure-risk assessment. In the process, we pave the ground-work towards incorporating auxiliary signals to improve our EEG-based seizure-risk assessment system. Our previous CNN results show an average AUC of 74.65% while we could achieve on an EEG-only BCNN an average AUC of 68.70%. This drop in performance is the cost of providing richer information to the patient at this stage of this research.**

## I. Introduction

There have been great interest recently in identifying biomarker for seizure susceptibility by looking into critical transitions in brain dynamics in order to enhance the precision of seizure forecasting in a cohort of patients with focal epilepsy [1]. These studies often require a very long recording that is not available and, in fact, are critically lacking. Chronic and often intracranial, electroencephalogram (EEG) recordings demonstrated some limited evidence of circadian, multidien, and circannual cycles in epileptic brain dynamics [2]. In any case of determining seizure-risk, we believe the certainty of our underlying decision-making algorithms must be communicated with patients or carers. In this paper, we introduce a probabilistic approach to seizure susceptibility assessment that transparently reveal uncertainty of our decision making regarding seizure forecasting.

The availability of a seizure forecasting system that can notify patients or their carers about forthcoming seizure-risk can drastically improve the quality of life for patients and also the chance to develop innovative interventions and preventative therapies. There has been an increasing number of studies on forecasting seizures; many of them use the signal-modal approach that is based on electroencephalogram (EEG) signals. However, there are several ways to utilize multi-modal signals to improve the performance of the single-modal approach. [3] For example, combined time and frequency features from EEG and ECG to perform seizure prediction with support vector machine (SVM). This method considers ECG as additional features and treats them the same way with features extracted from EEG. Another approach leverages circadian rhythms of seizure occurrences extracted from EPILEPSIA scalp-EEG dataset, as a post-processing technique to improve the seizure forecasting performance. The post-processing is independent of the training process of the forecasting system.

In this work, we use probabilistic programming and propose a framework to incorporate other relevant information into an EEG-based seizure forecasting system. As an advantage of using probabilistic programming, our system is able not only to forecast upcoming seizures but also to quantify the uncertainty level of its decision making.

## II. Method

### A. Bayesian convolutional neural network

In this paper, we will use variational inference to approximate posterior densities for Bayesian models [4]. Consider $x = x_{1:n}$ as a set of observed variables and $z = z_{1:m}$ as a set of hidden variables, with joint density $p(z, x)$. The inference problem is to calculate the conditional density of the hidden variables given the observed variables, $p(z|x)$.

$$p(z|x) = \frac{p(z,x)}{p(x)} , \tag{1}$$

where $p(x)$ is intractable in many models [4].

Variational inference overcomes this by specifying a variational family $\mathscr{Q}$ over the hidden variables. The inference problem becomes finding the best candidate $q(z) \in \mathscr{Q}$ that is closest in Kullback-Leibler (KL) divergence to $p(z|x)$. The optimization subsequently can be achieved by maximizing a function called the

* Corresponding author.

N.D. Truong, Y. Yang, C. Maher, and O. Kavehei are with the School of Biomedical Engineering, Faculty of Engineering, The University of Sydney, NSW 2006, Australia.

A. Nikpour is with Comprehensive Epilepsy Service and Department of Neurology at the Royal Prince Alfred Hospital, NSW 2050, and Sydney Medical School, The University of Sydney, NSW 2006, Australia.

E-mail: {duy.truong, yikai.yang, christina.maher, omid.kavehei} @sydney.edu.au, armin@sydneyneurology.com.au.

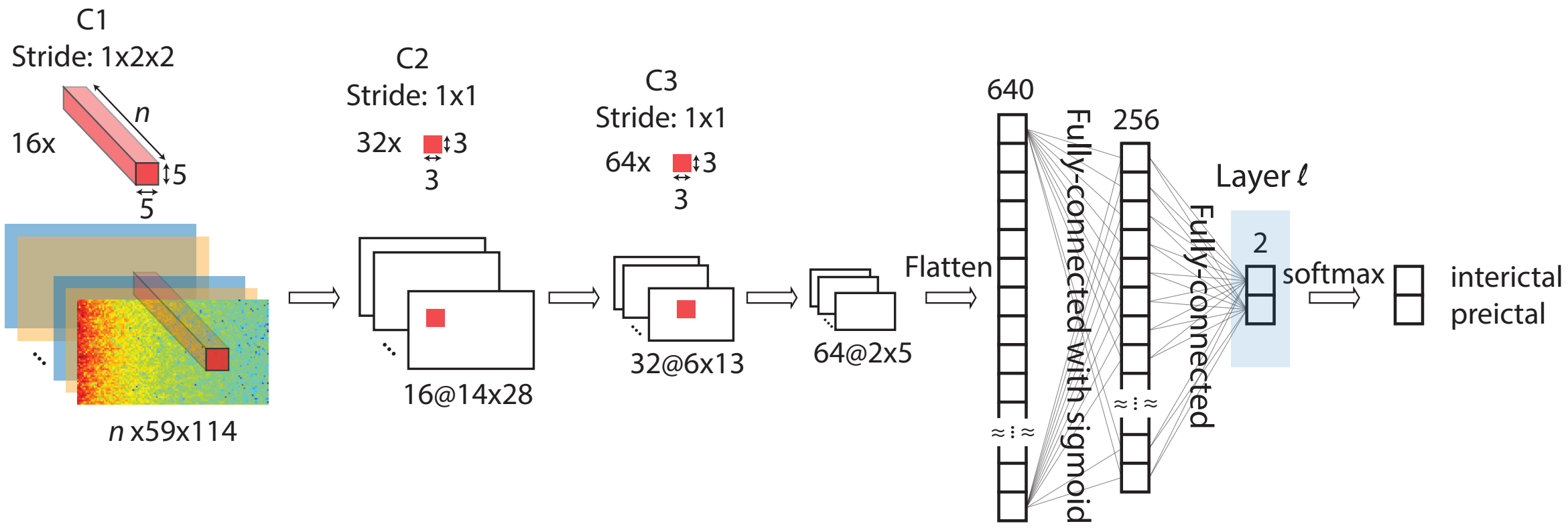

Fig. 1: Architecture of Bayesian convolutional neural network (BCNN). Different from a conventional convolutional neural network, each trainable parameter of a BCNN is a distribution. In our work, we employ normal distribution with mean value and standard deviation are trainable parameters. Layer $l$ is chosen to incorporate other relevant data for seizure forecasting by using Eqn. 8.

TABLE I: The EPILEPSIAE scalp-EEG dataset.

| Patient | Gender | Age | No. of seizures | No. of leading seizures* | Interictal hours |
|---|---|---|---|---|---|
| Pat1 | male | 36 | 11 | 11 | 68.9 |
| Pat2 | female | 46 | 8 | 8 | 114.9 |
| Pat3 | male | 41 | 8 | 8 | 96.3 |
| Pat4 | female | 67 | 5 | 5 | 126 |
| Pat5 | female | 52 | 8 | 8 | 204.1 |
| Pat6 | male | 65 | 8 | 7 | 92.2 |
| Pat7 | male | 36 | 5 | 5 | 75.7 |
| Pat8 | male | 26 | 22 | 11 | 65.6 |
| Pat9 | male | 47 | 6 | 6 | 51.1 |
| Pat10 | male | 44 | 11 | 11 | 60.7 |
| Pat11 | male | 48 | 14 | 14 | 57.8 |
| Pat12 | male | 28 | 9 | 9 | 94.1 |
| Pat13 | male | 46 | 8 | 8 | 101.3 |
| Pat14 | female | 62 | 6 | 6 | 115.7 |
| Pat15 | female | 41 | 5 | 5 | 82.8 |
| Pat16 | female | 15 | 6 | 6 | 51.1 |
| Pat17 | female | 17 | 9 | 9 | 82.4 |
| Pat18 | male | 47 | 7 | 6 | 133 |
| Pat19 | male | 32 | 22 | 21 | 75.4 |
| Pat20 | male | 47 | 7 | 7 | 115.3 |
| Pat21 | female | 31 | 8 | 8 | 106.6 |
| Pat22 | male | 38 | 7 | 7 | 88.2 |
| Pat23 | male | 50 | 9 | 9 | 179.6 |
| Pat24 | female | 54 | 10 | 10 | 36.2 |
| Pat25 | male | 42 | 8 | 8 | 109.8 |
| Pat26 | male | 13 | 9 | 9 | 97.1 |
| Pat27 | male | 58 | 9 | 8 | 99.9 |
| Pat28 | female | 35 | 9 | 9 | 95.2 |
| Pat29 | male | 50 | 10 | 10 | 111.9 |
| Pat30 | female | 16 | 12 | 12 | 92.5 |

* We are considering leading seizures only. Seizures that are less than 30 min away from the previous one are considered as one seizure only and the onset of leading seizure is used as the onset of the combined seizure.

evidence lower bound (ELBO) which is equivalent to minimizing the KL divergence between $q(z)$ and $p(z|x)$. ELBO is expressed as follows [4]:

$$\begin{aligned}\mathrm{ELBO}(q) &= \mathbb{E}\big[\log p(z,x)\big] - \mathbb{E}\big[q(z)\big] \\ &= \mathbb{E}\big[\log p(x|z)\big] + \mathbb{E}\big[\log p(z)\big] - \mathbb{E}\big[q(z)\big] \\ &= \mathbb{E}\big[\log p(x|z)\big] - \mathrm{KL}\big(q(z)\|p(z)\big)\end{aligned} \tag{2}$$

Stochastic variational inference was proposed in [5] to help Bayesian neural networks scale efficiently to large datasets. Particularly, this method generates noisy estimates of the natural gradient of the ELBO by repeatedly subsampling (mini-batch) the dataset. The loss function can be defined as the negative of ELBO; i.e., minimizing the loss is equivalent to maximizing the ELBO.

$$\mathrm{loss} = -\mathrm{ELBO}(q) = -\mathbb{E}\big[\log p(x|z)\big] + \mathrm{KL}\big(q(z)\|p(z)\big) \tag{3}$$

In an EEG-based seizure prediction system, $x$ is the EEG signals, and $z$ is a variable indicating a seizure to occur in the time window $\mathscr{T} = [\mathrm{SPH}:\mathrm{SPH}+\mathrm{SOP}]$.

### *B. Probabilistic convolutional neural network with data fusion*

In this section, we will incorporate signals other than EEG signals into the Bayesian CNN. We want to estimate the probability of having seizure given EEG signals, $p(z|x)$, which is the output of the Bayesian CNN. Besides EEG signals, we have other relevant data and want to combine all the information for seizure forecasting. Let us start with EEG signals and one other signal called $d$. Using Bayes theorem, the posterior probability of having a seizure in the time-window $\mathscr{T}$ can be expressed as:

$$p(z|x,d) = \frac{p(d|z,x)\,p(z|x)}{p(d|x)} \tag{4}$$

Assume $x$ and $d$ are independent, we can rewrite (4) as follows.

$$p(z|x,d) = \frac{p(d|z)\,p(z|x)}{p(d)} \tag{5}$$

Similarly, for two extra signals, $d_1$ and $d_2$ with an assumption that $x$, $d_1$ and $d_2$ are independent of each other, the posterior probability of having seizure in the time window $\mathscr{T}$ can be expressed as:

$$\begin{aligned}p(z|x,d_1,d_2) &= \frac{p(d_1|z,d_2)\,p(z|x,d_2)}{p(d_1|d_2)} \\ &= \frac{p(d_1|z)\,p(z|x,d_2)}{p(d_1)}\end{aligned} \tag{6}$$

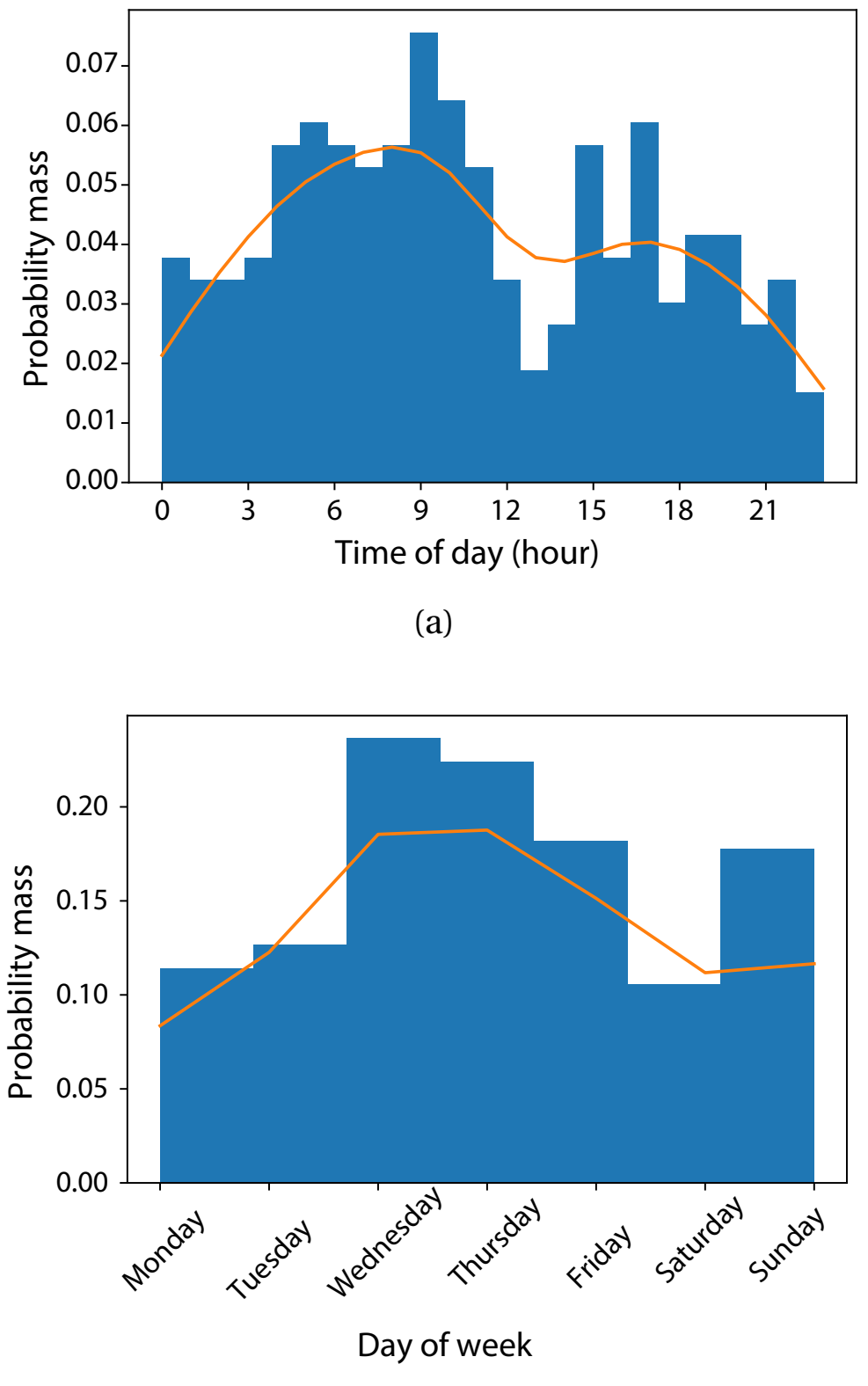

Fig. 2: Probability distribution of (a) time-of-day (ToD) and (b) day-of-week (DoW) of seizure occurrences in the EPILEPSIAE scalp-EEG dataset.

Substitute Eqn. (5) (with $d$ replaced by $d_2$) to Eqn. (6), we have:

$$p(z|x, d_1, d_2) = \frac{p(d_1|z)p(d_2|z)p(z|x)}{p(d_1)p(d_2)} \quad (7)$$

To estimate $p(d_1|z)$ and $p(d_2|z)$, we applied a kernel density estimation using Gaussian kernels on a histogram containing time-of-day (ToD) and day-of-week (DoW) of seizure occurrences (see Fig. 2) [6]. Note that here we approximate $p(d_1|z) \approx p(d_1|z')$ and $p(d_2|z) \approx p(d_2|z')$, where $z'$ is the variable indicating an occurrence of seizure. The approximation is reasonable because we choose the time window $\mathscr{T} = [5\text{ min} : 35\text{ min}]$ which is less than one hour.

To incorporate Eqn. 7 into the training of the Bayesian CNN, we modify the output of last fully-connected layer, $l$, before softmax activation as follows.

$$\text{new-output}_l = \frac{p(d_1|z')p(d_2|z') \times \text{output}_l}{p(d_1)p(d_2)}, \quad (8)$$

where $p(d_1|z')$ and $p(d_2|z')$ are derived from the kernel density estimation (see Fig. 2) and $p(d_1) = 1/24$, $p(d_2) = 1/7$. Note that Eqn. 7 and Eqn. 8 can be extended with more extra signals $d$ given that they are independent on each other.

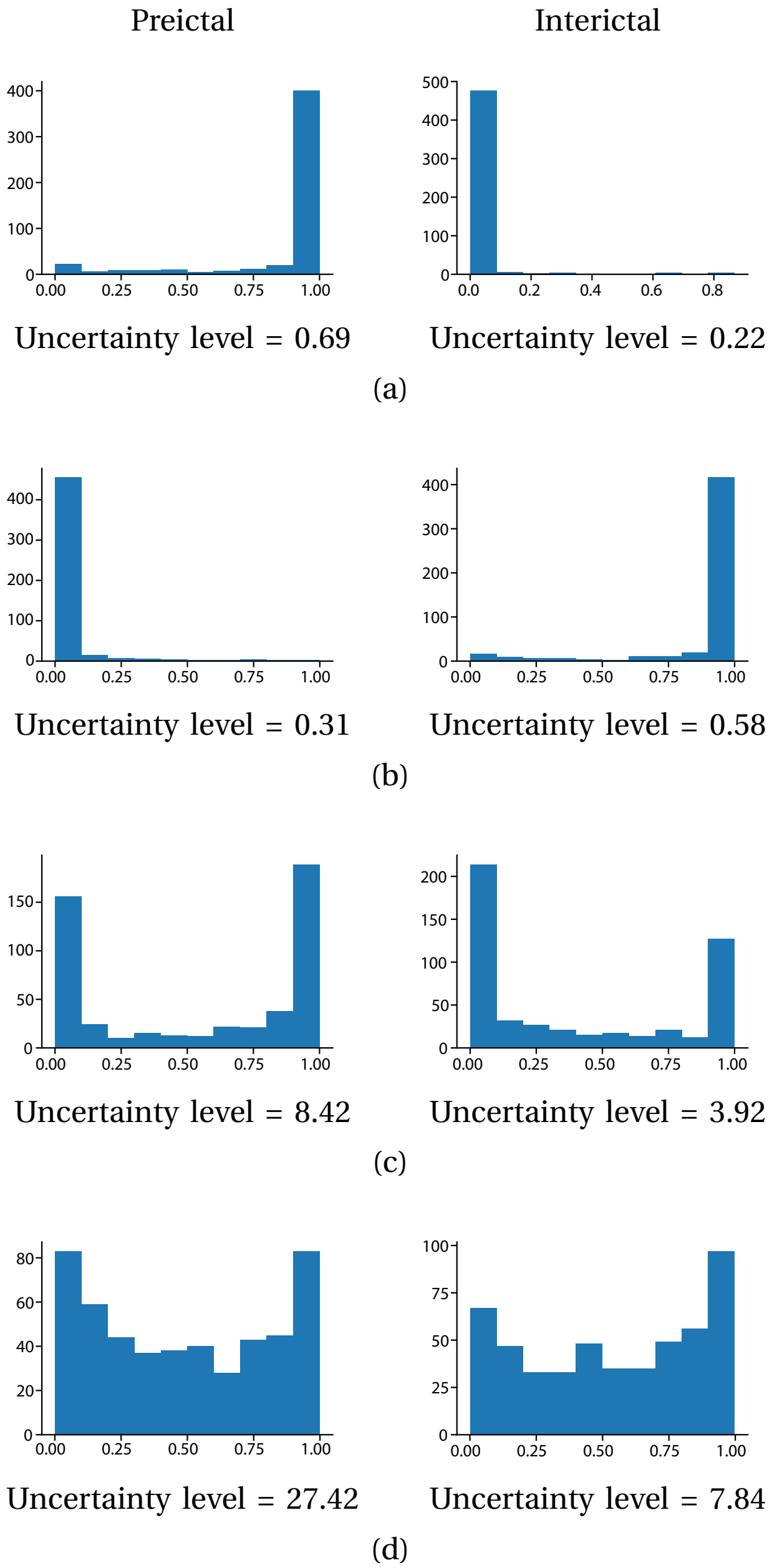

Fig. 3: Inference values by sampling the output of Bayesian convolutional neural network 500 times. Left: preictal; Right: interictal. (a) Correct predictions. (b) Wrong predictions. (c) Poor predictions with high standard deviation. (d) Poor predictions with low standard deviation.

## III. Dataset

EPILEPSIAE is the largest epilepsy database that contains EEG data from 275 patients [7]. In this work, we analyze scalp-EEG of 30 patients with 261 leading seizures and 2881.4 interictal hours in total. The time-series EEG signals were recorded at a sampling rate of 256 Hz and from 19 electrodes. Seizure onset information obtained by two methods, namely EEG based and video analysis, is provided. In our study, we use seizure onset information using an EEG based technique, where the onsets were determined by visual

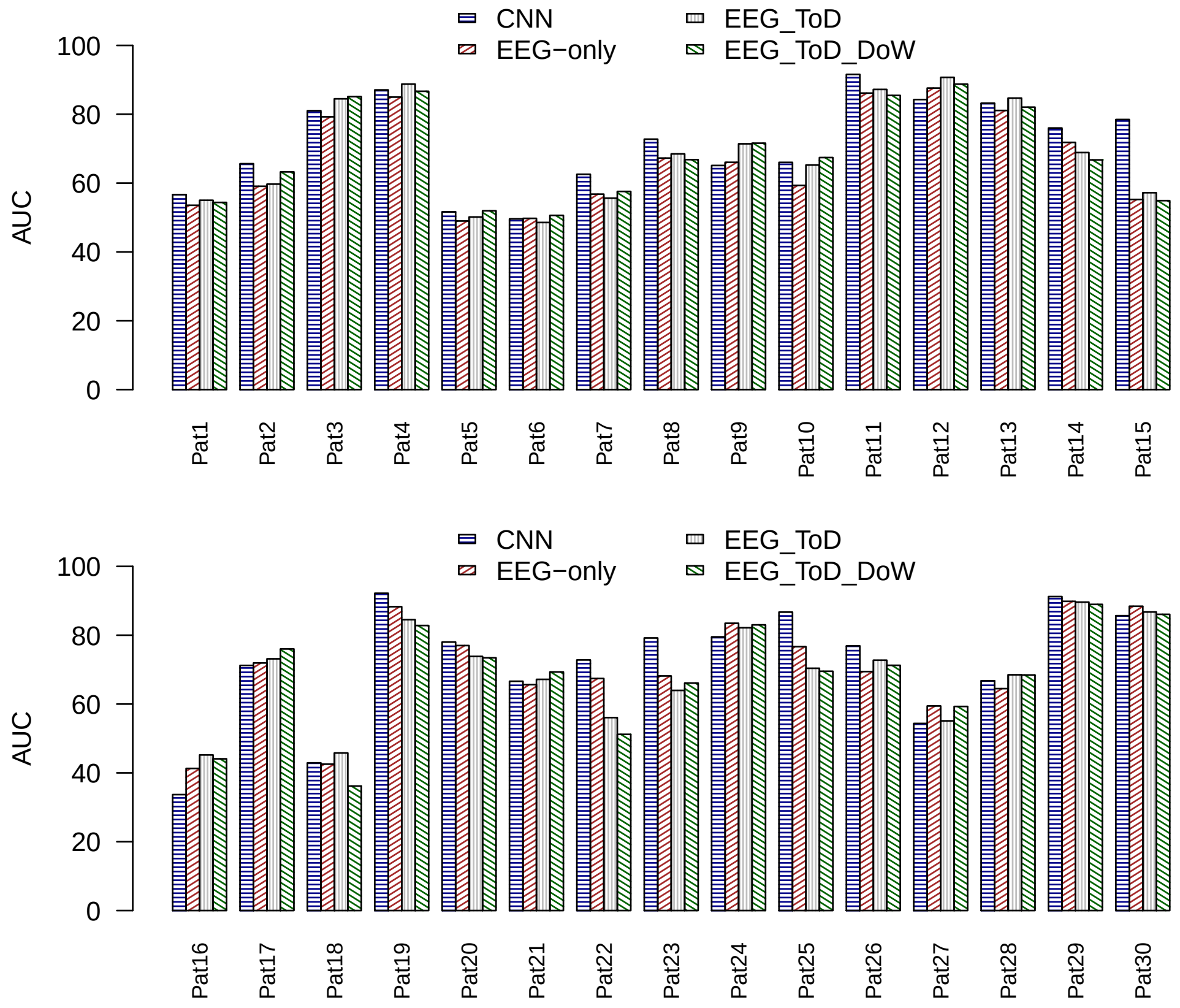

Fig. 4: Seizure prediction performance using Bayesian convolutional neural network (BCNN). CNN: Convolutional neural network, average AUC is 74.65%; EEG-only: BCNN using EEG signals only, average AUC is 68.70%; EEG_ToD: BCNN using EEG signals and time-of-day (ToD), average AUC is 69.03%; EEG_ToD_DoW: BCNN using EEG signals, time-of-day and day-of-week (DoW), average AUC is 68.63%. This information are used for post-processing, independent of the training process of the forecasting system.

inspection of EEG signals performed by an experienced clinician [7].

## IV. Results

We test the Bayesian convolutional neural network (BCNN) with the EPILEPSIAE EEG dataset with and without other signals: time-of-day and day-of-week. We also compare a seizure prediction system using a convolutional neural network (CNN) as a baseline. Compared to CNN, BCNN has around 6% lower in AUC. By using the time-of-day information, the overall performance of BCNN is slightly improved by 0.3%. However, using both time-of-day and day-of-week does not bring gain in performance.

## V. Discussion

Bayes convolutional neural network (BCNN) is able to generate the distribution of its output for each input. We sampled the output of the BCNN by feeding forward the same input through the BCNN 500 times. Fig. 3 depicts typical sampled distributions of a BCNN's output. Fig. 3(a) and (b) show correct predictions and wrong predictions with low uncertainty. Fig. 3(c) and (d) both show high uncertainty but one with high standard deviation (c), the other with low standard deviation (d). We quantify the uncertainty level of the BCNN's decision making with Eqn. 9 below. The numerator takes into account the variability of the output with the standard deviation (std). The denominator considers the case where the output has a uniform-like distribution. Uncertainty levels of different types of prediction distributions are illustrated in Fig. 3.

$$\text{Uncertainty level} = \frac{\text{std}_{\text{inference values}}}{\left|\text{mean}_{\text{inference values}} - 0.5\right|} \tag{9}$$

Though we trained the BCNN with two types of EEG signals: preictal - 35 to 5 minutes before seizure onset, and interictal - at least 4 hours away from any seizures, we are interested in how the BCNN performs with continuous EEG recording. We ran inference over thirteen hours of continuous EEG recording for one of the best performers, Patient 4, that consisted of two seizures. In Fig. 5, we plot both the prediction scores (from 0 to 1, where higher values indicate a higher probability of having a seizure) and the corresponding uncertainty levels of the BCNN. The prediction scores get higher values when it is closer to the first seizure onset. It is interesting to note that at around time 40 and 60 minutes, there are two predictions with high scores, but the uncertainty levels were also high, which means the BCNN "thinks" that there might be a seizure incoming, but the BCNN has a very low confidence about its decision. In the period from about one hour to seizure onset, we can see prediction scores were mostly high, but there were many low prediction scores with high uncertainty levels. We suggest that the "patterns" related to seizure prediction only occur at a particular time rather than consistently the whole preictal duration.

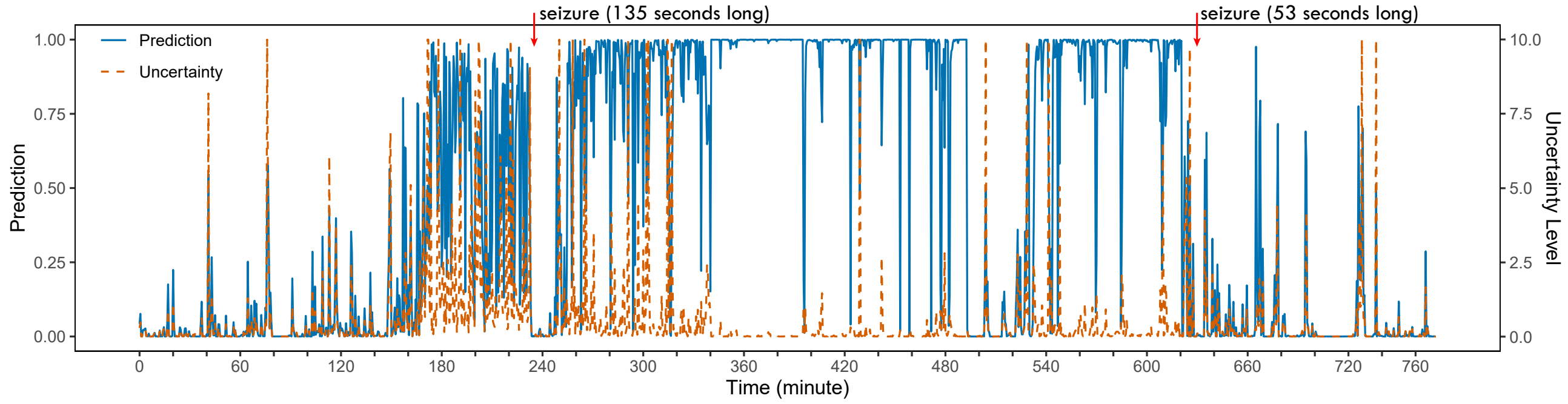

Fig. 5: Prediction score and uncertainty level produced by Bayesian convolution neural network in seizure forecasting task over thirteen hours of continuous EEG recording. For the sake of visualization, for all uncertainty levels higher than 10, we set them to 10.

## VI. Conclusion

Epileptic seizure forecasting is still a substantially challenging task with a consequential impact on patients' quality of life and their caregivers. While some patient-specific demonstrated excellent performance in a subset of patients, generalized predictions on non-invasive EEG recordings with the capability to work well on most patients, have been a great challenge. This work presented an innovative approach to incorporate uncertainty and auxiliary signals (data-fusion) information in seizure-risk forecasting. These informative warning signals will be invaluable for patient or caregiver decision making in employing any risk-mitigation intervention or therapies. We built our method based on the Bayesian convolutional neural network to provide an insight into the uncertainty level of seizure-risk prediction.

Further investigation is required to highlight the value of a systematic approach in incorporating vital signals, such as heart rate variability, blood oxygen level (SpO2), and temperature, as well as validated circadian and multidien cycles information in seizure forecasting in patients living with epilepsy.

## Acknowledgment

This research is funded by a Sydney Research Accelerator (SOAR) Fellowship at The University of Sydney.